\documentclass{revtex4}

\usepackage{amsthm,amssymb,amsmath}				
\usepackage{graphicx,comment}
\usepackage{float}   
\usepackage{xcolor}

\begin{document}

\title{KG-particles in a cosmic string rainbow gravity spacetime in mixed magnetic fields}
\author{Omar Mustafa}
\email{omar.mustafa@emu.edu.tr}
\affiliation{Department of Physics, Eastern Mediterranean University, 99628, G. Magusa, north
Cyprus, Mersin 10 - Turkiye.}

\begin{abstract}
\textbf{Abstract:}\ {We investigate and report the effects of rainbow gravity on the spectroscopic structure of KG-oscillators in a mixed magnetic field (in the sense that it has the usually described as a uniform and a non-uniform magnetic fields, each at a time) introduced by the 4-vector potential  $A_\mu=(0,0,A_\varphi,0)$, where $A_\varphi=B_1 r^2/2+B_2 r$, and $B_1$ and $B_2$ are magnetic field strengths. We also discuss and report the effects of such a mixed magnetic field on the spectra of KG-oscillators in cosmic string rainbow gravity. In so doing, we introduce a new and quite handy \textit{conditionally exact solution} associated with the truncation of the biconfluent Heun functions into polynomials. Using a loop quantum gravity motivated rainbow functions, we observe interesting effects when the magnetic field strength $B_2$ grows up from zero. Such effects include energy levels crossings which, in this case, turns the spectra of the KG-oscillators upside down. Moreover, Landau-like signature on the spectra are observed and discussed.  Yet, interestingly, we also observe that rainbow gravity affects the magnetic field as well, in the sense that the magnetic field becomes probe particle energy-dependent one. }


\end{abstract}

\maketitle

\section{Introduction}
Topological defects in the spacetime fabric are stable configurations of matter that are believed to be formed during the rapid expansion and cooling process at the phase transitions in the very early universe  \cite{CR1,CR2}. The type of the topological defect is determined by the symmetry properties of the matter and the nature of the phase transition. Domain walls \cite{CR020,CR021} (two-dimensional objects formed when a discrete symmetry is broken), cosmic strings \cite%
{CR021,CR3,CR4,CR5,CR6} (one-dimensional objects formed when cylindrical symmetry is broken), and global monopoles (GM) \cite{CR2,CR3,CR6,CR7} (zero-dimensional objects formed when spherical symmetry is broken), are among such feasible  topological defects, to mention a few. Cosmic strings are the topological defects of our interest in the current study. 

Rainbow gravity (RG), on the other hand, has inspired research studies for years \cite{CR8,CR9,CR10,CR11,CR12},  for being a semi classical extension of the doubly special relativity. In the ultra high energy regime (i.e., ultraviolet limit) , nevertheless, RG suggests that the energy of the probe particle affects the spacetime background, in the sense that the corresponding spacetime metric (using the units $c=\hbar=8\pi G=1$) takes the energy-dependent form  \cite{CR8,CR9,CR10,CR11,CR12,CR13,CR14,CR15,CR16,CR17,CR18,CR19,CR20,CR21,CR22,CR23,CR24,CR25,CR26,CR27}%
\begin{equation}
ds^{2}=-\frac{1}{g_{_{0}}\left( y\right) ^{2}}dt^{2}+\frac{1}{g_{_{1}}\left(
y\right) ^{2}}\left( dr^{2}+\alpha ^{2}\,r^{2}d\varphi ^{2}+dz^{2}\right);\,\,0\leq \left( y=E/E_{p}\right) \leq 1.
\label{a1}
\end{equation}%
Where $0<\alpha^2=1-\eta/2\pi<1 $ is a constant related to the deficit angle of the conical spacetime,  $G$ is the Newton's constant,  $\eta $ is the linear mass density of the cosmic string, and  $g_{_{k}}(y) , k=0,1$, are called the rainbow functions. The corresponding metric tensor $g_{\mu \nu }$ is given by%
\begin{equation}
g_{\mu \nu }=diag\left( -\frac{1}{g_{_{0}}\left( y\right) ^{2}},\frac{1}{%
g_{_{1}}\left( y\right) ^{2}},\frac{\alpha ^{2}\,r^{2}}{g_{_{1}}\left(
y\right) ^{2}},\frac{1}{g_{_{1}}\left( y\right) ^{2}}\right) ;\;\mu ,\nu
=t,r,\varphi ,z,  \label{a11}
\end{equation}%
with 
\begin{equation}
\det \left( g_{\mu \nu }\right) =-\frac{\alpha ^{2}\,r^{2}}{g_{_{0}}\left(
y\right) ^{2}g_{_{1}}\left( y\right) ^{6}}\Longrightarrow g^{\mu \nu
}=diag\left( -g_{_{0}}\left( y\right) ^{2},g_{_{1}}\left( y\right) ^{2},%
\frac{g_{_{1}}\left( y\right) ^{2}}{\alpha ^{2}\,r^{2}},g_{_{1}}\left(
y\right) ^{2}\right) .  \label{a12}
\end{equation}%
Under RG model,  however, the Planck energy $E_{p}$ represents a threshold that separates quantum from classical mechanical descriptions and introduces itself as another invariant energy scale, in addition to the speed of light. RG justifies, therefore, the modified relativistic energy-momentum dispersion relation%
\begin{equation}
E^{2}g_{_{0}}\left( y\right) ^{2}-p^{2}c^{2}g_{_{1}}\left( y\right)
^{2}=m^{2}c^{4};\;0\leq \left( y=E/E_{p}\right) \leq 1,  \label{a2}
\end{equation}%
where $mc^{2}$ is its rest mass energy. Such a modification is significant in the high energy regime and is restricted to retrieve the standard general relativity dispersion relation in the infrared limit so that%
\begin{equation}
\lim\limits_{y\rightarrow 0}g_{_{k}}\left( y\right) =1;\;k=0,1.  \label{a3}
\end{equation}%
At this point, one should observe that the rainbow function variable $y$ satisfies $0\leq \left( y=E/E_{p}\right) \leq 1$  This would, in turn, suggest that as long as relativistic quantum particles ($E=+|E|=E_+$) and anti-particles 
($E=-|E|=E_-$) are in point, it is mandatory to use a fine tuned rainbow function variable $y=|E|/E_p$. Using such a fine tuned rainbow function variable, we have very recently shown \cite{CR28,CR29} that rainbow gravity works on relativistic particles and anti-particles alike. Otherwise, rainbow gravity shows its effect on particles only and anti-particles would have indefinitely unbounded energies. Violating, therefore, the intended characteristic of rainbow gravity that the Planck energy $E_{p}$ is the maximum possible energy for particles and anti-particles, i.e., $|E|=E_p$, as observed in \cite{CR16,CR17,CR19,CR20,CR21,CR27}, to mention a few. We have found \cite{CR28,CR29} that the loop quantum gravity motivated \cite{CR30,CR31} fine tuned rainbow functions pair%
\begin{equation}
    g_0(y)=1,\,\, g_1(y)=\sqrt{1-\epsilon\, y^n}=\sqrt{1-\epsilon \left(\frac{|E|}{E_p}\right)^n};\,\, n=1,2, \label{a31}
\end{equation}%
completely complies with the rainbow gravity model and secures Planck energy $E_{p}$ as the maximum energy for particles and anti-particles. Whereas, the rainbow function pairs $g_0(y)=g_1(y)=(1-\epsilon y)^{-1}$ used in resolving the horizon problem \cite{CR22,CR32}, and  $g_0(y)=(e^{\epsilon y}-1)/\epsilon y$, $g_1(y)=1$ obtained from the spectra of gamma-ray bursts \cite{CR13} have only partially complied with the intended rainbow gravity effect. Hereby, we shall use the loop quantum gravity motivated pairs in (\ref{a31}) in the current study.

Yet, the inclusion of a 4-vector potential $A_\mu=(0.0,A_\varphi,0)$ \cite{CR33}, with $A_\varphi=A_{1,\varphi}+A_{2,\varphi}=B_1 r^2/2+B_2 r$ would introduce a magnetic field $\textbf{B}=B_z \hat{z}$, where%
\begin{equation}
    B_z^2=\frac{1}{2}F_{\mu\nu}F^{\mu\nu}=g^{rr}g^{\varphi\varphi}(\partial_r A_\varphi-\partial_\varphi A_r)^2\Rightarrow B_z=\frac{g_1(y)^2}{\alpha}\left[B_1+\frac{B_2}{r}\right]. \label{a32}
\end{equation}%
Obviously, in no rainbow gravity (i.e., $\epsilon=0\rightarrow g_1(y)=1$),  the first term would identify $B_1$ as the strength of the uniform magnetic field part, and the second term would identify $B_2$ as the strength of the non-uniform (i.e., $r$-dependent) magnetic field part. Yet, $B_2$ can be identified as the strength of a uniform magnetic field (i.e., constant) at every specific value of $r$. The magnetic fields produced by $A_{1,\varphi}$ and $A_{2,\varphi}$ are readily discussed in some details in \cite{CR33} (the reader is advised to refer to for more details). However, under rainbow gravity, we observe that both parts of the magnetic field in (\ref{a32}) are dependent on the energy of the probe particle/anti-particle. The classifications of uniform and/or non-uniform magnetic fields is a matter of perspective. We shall adopt the simple notion of a \textit{mixed magnetic field} hereinafter, therefore.

Rainbow gravity, nevertheless, has been a subject of different studies. To mention a few,  thermodynamical properties of black holes in RG  \cite{CR33.1,CR33.2,CR33.3,CR33.4,CR33.5,CR33.6,CR33.7,CR33.8}, stability of neutron stars \cite{CR33.9}, TeV photons \cite{CR25}, and $f(R)$ theories \cite{CR33.10}. On the relativistic and non-relativistic quantum mechanical side, moreover, studies are carried out, for example, on Landau levels \cite{CR17}, Aharonov-Bohm effects on Dirac oscillators \cite{CR33.11,CR33.12}, KG-Coulombic particles \cite{CR27,CR28}, DKP-particles by Hosseinpour et al. \cite{CR12}, etc.

In the current methodical proposal, we shall investigate KG-particles in a cosmic string rainbow gravity spacetime in a mixed magnetic fields. In section 2, we start with KG-particles in cosmic string fine tuned rainbow gravity spacetime and in a mixed magnetic field given by (\ref{a32}). In the same section, we consider the KG-oscillators and bring the corresponding KG equation into the one-dimensional Schr\"{o}dinger oscillator form. The resulting Schr\"{o}dinger form consequently includes a harmonic oscillator, a Coulombic-like, and a linear type interaction potential terms. The solution of which is usually given in terms of the biconfluent Heun functions  $H_B(\alpha, \beta, \gamma, \delta, z)$ that are truncated to a polynomial of order $n_r\geq 0$  when $\gamma=2(n_r+1)+\alpha$. However, in the current methodical proposal, we shall truncate the biconfluent Heun functions to a polynomial of order $n_r+1$ and introduce a new condition that not only facilitates conditional exact solvability of the problem at hand but also retrieves  the condition $\gamma=2(n_r+1)+\alpha$. We do so in section 3. Therein, we report and discuss the most intriguing observation of the current study, represented by energy levels crossings that flip the spectra of the KG-oscillators upside down. This effect is a consequence of the structure of the mixed magnetic field used (and has nothings to do with RG).  In section 4 we discuss Landau-like signatures on the spectroscopic structure of the KG-oscillators. Our concluding remarks are given in section 5.

\section{KG-particles in cosmic string rainbow gravity spacetime and a mixed magnetic field}

In the cosmic string rainbow gravity spacetime background (\ref{a1}), a KG-particle of charge $e$ in the 4-vector potential $A_\mu=(0,0,A_\varphi,0)$, with $A_\varphi=B_1 r^2/2+B_2 r$  is described (in $c=\hbar =1$ units) by the KG-equation%
\begin{equation}
\frac{1}{\sqrt{-g}}\tilde{D}_{\mu }\left( \sqrt{-g}g^{\mu \nu }\tilde{D}%
_{\nu }\right) \Psi =m_\circ^{2}\Psi \Longrightarrow \frac{1}{\sqrt{-g}}\left(
D_{\mu }+\mathcal{F}_{\mu }\right) \sqrt{-g}g^{\mu \nu }\left( D_{\nu }-%
\mathcal{F}_{\nu }\right) \Psi =m_\circ^{2}\Psi .  \label{b1}
\end{equation}%
where $D_{\mu }$ is the gauge-covariant derivative given by $D_{\mu
}=\partial _{\mu }-ieA_{\mu }$ and admits minimal coupling, $m_\circ$ is the rest mass energy of the KG-particle, and $\mathcal{F}_\mu$ is usually used to incorporate KG-oscillators \cite{CR33.13,CR33.14}. This equation would, in a straightforward manner, imply, with%
\begin{equation}
\Psi \left( t,r,\varphi ,z\right) =\exp \left( i\left[ m \varphi
+kz-Et\right] \right) \phi \left( r\right) ,  \label{b2}
\end{equation}%
that%
\begin{equation}
\left\{ \tilde{E}^2+\left[ \partial _{r}^{2}+\frac{1}{r}\partial _{r}-M\left(
r\right) -\frac{1}{\alpha ^{2}\,r^{2}}\left( m-eA_{\varphi }\right) ^{2}\right] \right\} \phi \left( r\right) =0 ,  \label{b3}
\end{equation}%
where $m=0,\pm1,\pm2,\cdots$ is the magnetic quantum number,%
\begin{equation}
    \tilde{E}^2=\frac{g_0(y)^2 E^2-m_\circ^2-g_1(y)^2k^2}{g_1(y)}, \label{b4}
\end{equation}
and %
\begin{equation}
M\left( r\right) =\mathcal{F}_{r}^{\prime }+\frac{\mathcal{F}_{r}}{r}+%
\mathcal{F}_{r}^{2}, \label{b5}
\end{equation}%
We now substitute  $A_\varphi=B_1 r^2/2+B_2 r$, and $\mathcal{F}_r=\Omega r$  (to incorporate KG-oscillators \cite{CR33.13,CR33.14} in the process), to obtain%
\begin{equation}
\left\{ \partial _{r}^{2}+\frac{1}{r}\partial _{r}
-\frac{\tilde{m}^{2}}{r^{2}}-\tilde{\Omega}^{2}r^{2}+\frac{2\tilde{m}\tilde{B}_2}{r}-\tilde{B}_1\tilde{B}_2 r+\lambda^2 \right\} \phi
\left( r\right) =0,  \label{b6}
\end{equation}%
where%
\begin{equation}
\lambda^2 =\tilde{E}^2-(\tilde{B}_2^2-\tilde{m}\tilde{B}_1)-2\Omega\,;\; \tilde{\Omega}^2=\Omega^2+\frac{\tilde{B}_1^2}{4}\, ,\;\tilde{m}=\frac{m }{\alpha },\;\tilde{B}_i=\frac{%
eB_{i}}{\alpha }.  \label{b7}
\end{equation}%
Moreover, with $\phi \left( r\right) =U\left( r\right) /\sqrt{r}$ we obtain the two-dimensional radial KG-oscillators%
\begin{equation}
\left\{ \partial _{r}^{2}-\frac{\left( \tilde{m}^{2}-1/4\right) }{r^{2}}%
-\tilde{\Omega}^{2}r^{2}+\frac{2\tilde{m}\tilde{B}_2}{r}-\tilde{B}_1\tilde{B}_2 r+\lambda^2 \right\} U\left( r\right) =0.
\label{b8}
\end{equation}%
This equation is known to admit a solution in the form of biconfluent Heun functions so that%
\begin{equation}
    U(r)=r^{|\tilde{m}|+\frac{1}{2}}\, e^{\left(-\frac{\tilde{\Omega}r^2}{2}-\frac{\tilde{B}_1\tilde{B}_2 r}{2\tilde{\Omega}}\right)}\, H_B\left(2|\tilde{m}|,\frac{\tilde{B}_1\tilde{B}_2 }{\tilde{\Omega}^{3/2}}, \frac{\tilde{B}_1^2\tilde{B}_2^2+4\lambda^2\tilde{\Omega}^2}{4\tilde{\Omega}^3},-\frac{4|\tilde{m}|\tilde{B}_2}{\sqrt{\tilde{\Omega}}},\sqrt{\tilde{\Omega}} r\right). \label{b9}
\end{equation}%
The biconfluent Heun function $H_B(\alpha, \beta, \gamma, \delta, z)$ is truncated to a polynomial of order $n_r\geq 0$ for $\gamma=2(n_r+1)+\alpha$  to imply that%
\begin{equation}
    \lambda^2=2 \tilde{\Omega} (n_r+|\tilde{m}|+1)-\frac{\tilde{B}_1^2\tilde{B}_2^2}{4\tilde{\Omega}^2}. \label{b10}
\end{equation}%
This truncation condition is to be accompanied by additional conditions that facilitate a so called conditional exact solvability of the problem at hand. One of such additional conditions is discussed by Ronveaux \cite{CR34}, some other conditions on the confluent Heun functions are discussed by Ishkhanyan et al \cite{CR35}, and very recently a quit handy condition is introduced by Mustafa et al \cite{CR36}. The later is to be adopted and followed in the current methodical proposal to obtain a \textit{conditional exact solution}, as well as retrieve the results of some special examples, to be illustrated below . 

\section{Conditionally exact energies for KG-oscillators in cosmic string rainbow gravity spacetime and a mixed magnetic field}

We recollect equation (\ref{b6}) and use the substitution%
\begin{equation}
    \phi \left( r\right) =exp\left(-\frac{\tilde{\Omega}r^2}{2}-\frac{\tilde{B}_1\tilde{B}_2 r}{2\tilde{\Omega}}\right)\,\frac{R(r)}{\sqrt{r}}, \label{b11}
\end{equation}%
\begin{figure}[!ht]  
\centering
\includegraphics[width=0.3\textwidth]{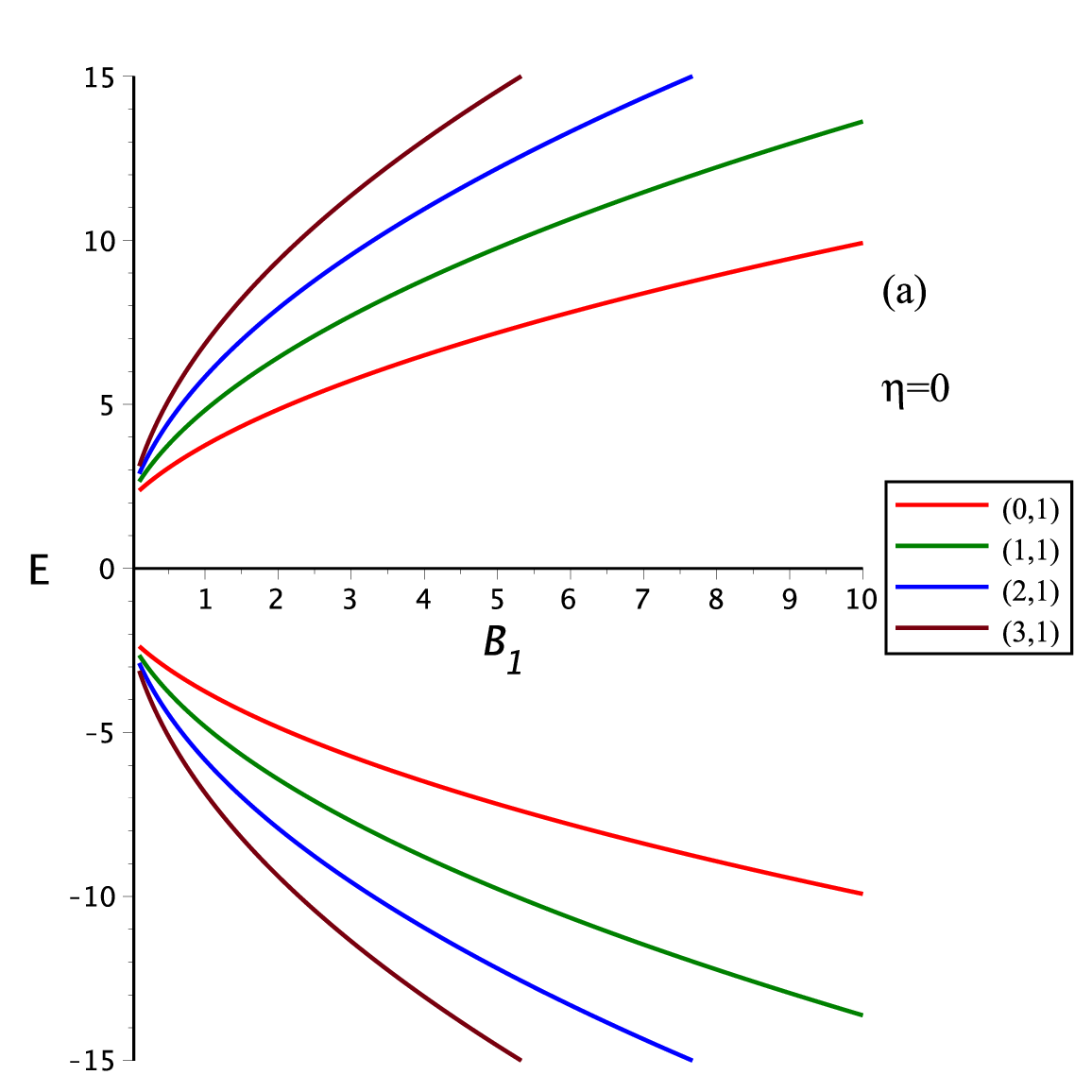}
\includegraphics[width=0.3\textwidth]{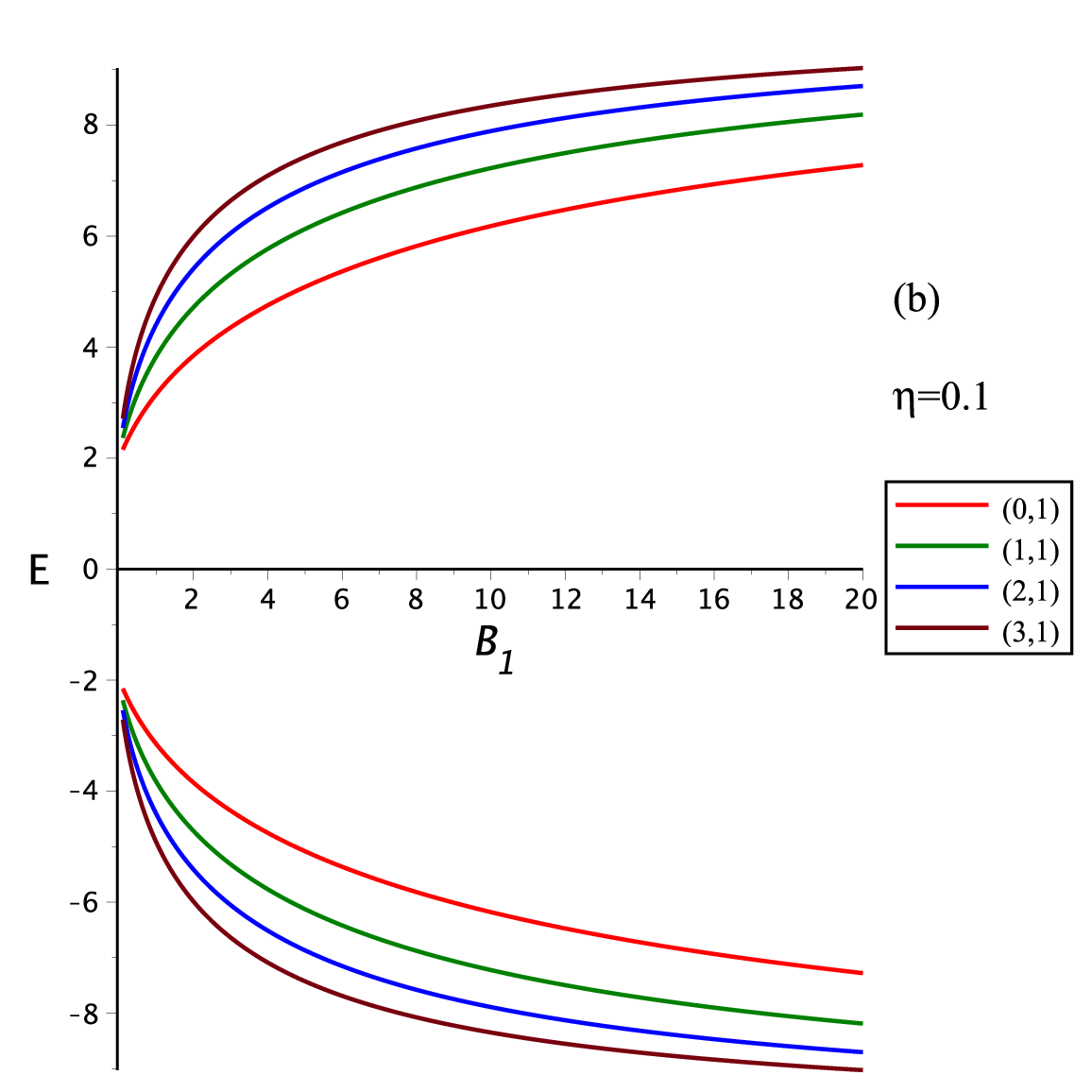} 
\includegraphics[width=0.3\textwidth]{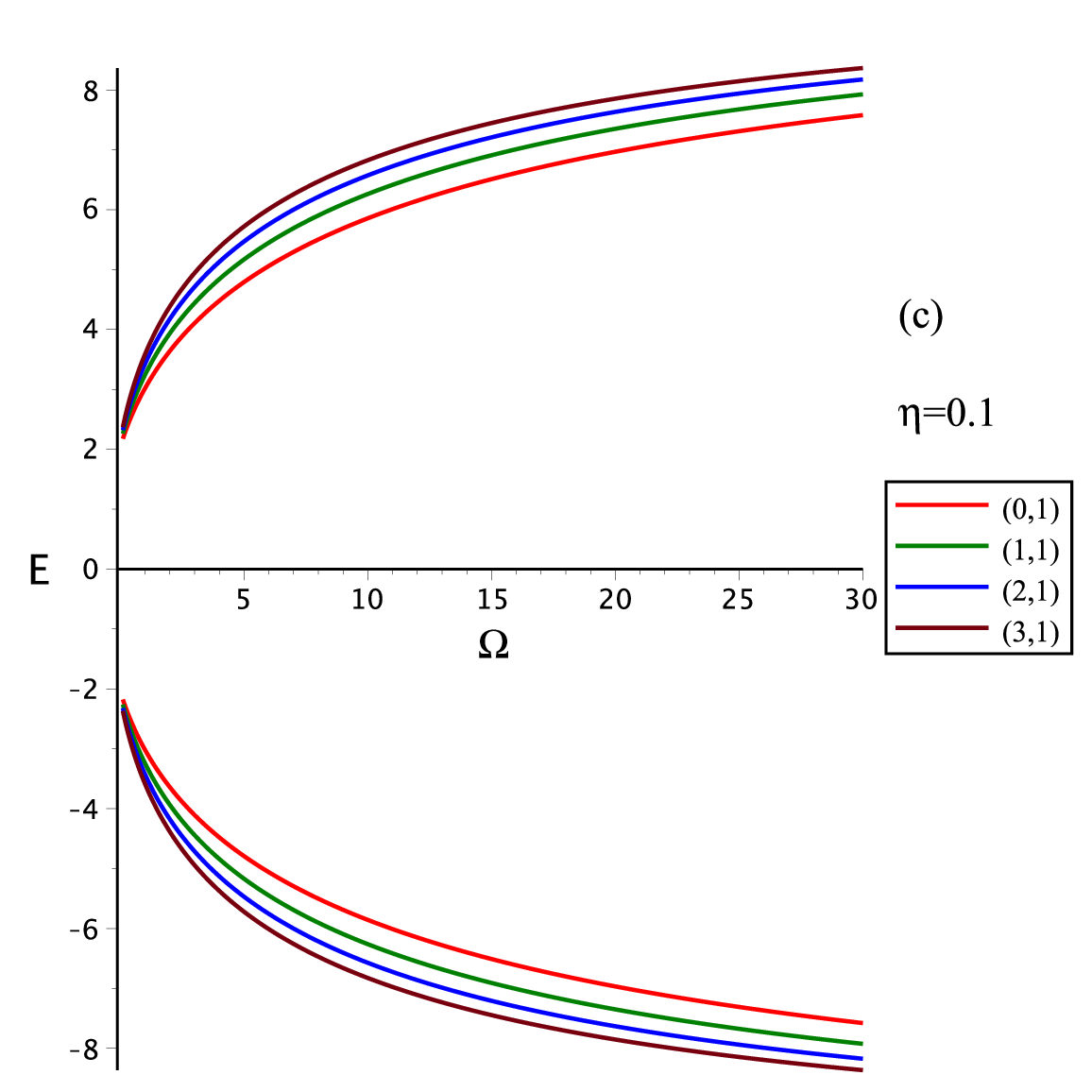}
\caption{\small 
{ The energy levels for the KG-oscillators in cosmic string rainbow gravity and a mixed magnetic field with the rainbow functions pair $g_0(y)=1,\, g_1(y)=\sqrt{1-\eta |E|}$, given by  Eq.(\ref{b27}),  plotted at $\alpha=0.5,\, m_\circ=1=q=k$  for $n_r=0,1,2,3,$ and $m=1$, where (a) $E$ against $B_1$ for $\eta=0$ (no rainbow gravity) and $B_2=1$, (b) $E$ against $B_1$ for $\eta=0.1$ and $B_2=1$, and (c) $E$ against the KG-oscillator frequency $\Omega$ for $\eta=0.1$ and $B_2=1$. }}
\label{fig1}
\end{figure}%
to obtain%
\begin{equation}
 -\frac{\tilde{\Omega}^2r^2}{2}R^{\prime\prime}(r)+\tilde{\Omega}r^2\left(\tilde{\Omega}^2r+ \frac{\tilde{B}_1\tilde{B}_2}{2}\right)R^\prime(r)-\frac{1}{2}(P_1 r^2+P_2 r-P_3)R(r)=0, \label{b12}
\end{equation}%
where%
\begin{equation}
    P_1=\lambda^2\tilde{\Omega}^2+\frac{\tilde{B}_1^2\tilde{B}_2^2}{4}-\tilde{\Omega}^3\,,\,\, P_2=2\tilde{m}\tilde{B}_2\tilde{\Omega}^2\,,\,\, P_3=\tilde{\Omega}^2\left(\tilde{m}^2-\frac{1}{4}\right). \label{b13}
\end{equation}%
We now use a power series expansion in the form of%
\begin{equation}
    R(r)=r^\nu \sum\limits_{j=0}^{\infty }A_j\, r^j, \label{b14}
\end{equation}%
to obtain%
\begin{gather}
\sum\limits_{j=0}^{\infty }\left\{ A_j\,\left[\tilde{\Omega}^3(j+\nu)-\frac{P_1}{2}\right] 
+A_{j+1}\left[\frac{\tilde{B}_1\tilde{B}_2 \tilde{\Omega}}{2}\left( j+\nu +1\right) -\frac{P_2}{2}\right] \right.   \notag \\
\left. 
+A_{j+2}\left[ \frac{P_3}{2}-\frac{\tilde{\Omega}^2}{2}\left( j+\nu +1\right)
\left( j+\nu +2\right)\right] \right\} r^{j+\nu +2} \notag \\
+\left\{ A_{0}\left[\frac{\tilde{B}_1\tilde{B}_2 \tilde{\Omega}\nu}{2}  -\frac{P_2}{2}\right] +A_{1}\left[ \frac{P_3}{2}-\frac{\tilde{\Omega}^2}{2}\nu \left( \nu +1\right)
\right] \right\} r^{\nu+1}  \notag \\
+A_{0}\left[ \frac{P_3}{2}-\frac{\tilde{\Omega}^2}{2}\nu \left( \nu -1\right)
\right] r^{\nu}=0.
\label{b15}
\end{gather}%
Which would suggest that since $A_0\neq 0$ we get%
\begin{equation}
   \frac{P_3}{2}-\frac{\tilde{\Omega}^2}{2}\nu \left( \nu -1\right)=0 \Rightarrow \nu=|\tilde{m}|+\frac{1}{2}, \label{b16}
\end{equation}%
\begin{equation}
    A_{0}\left[\frac{\tilde{B}_1\tilde{B}_2 \tilde{\Omega}\nu}{2}  -\frac{P_2}{2}\right] +A_{1}\left[ \frac{P_3}{2}-\frac{\tilde{\Omega}^2}{2}\nu \left( \nu +1\right)
\right]=0 \Rightarrow A_{1}=\frac{\tilde{B}_2\left[\frac{\tilde{B}_1}{2}\left(|\tilde{m}|+\frac{1}{2}\right)-\tilde{m}\tilde{\Omega}\right]}{\tilde{\Omega}\left(|\tilde{m}|+\frac{1}{2}\right)}A_0, \label{b17}
\end{equation}%
and%
\begin{figure}[!ht]  
\centering
\includegraphics[width=0.35\textwidth]{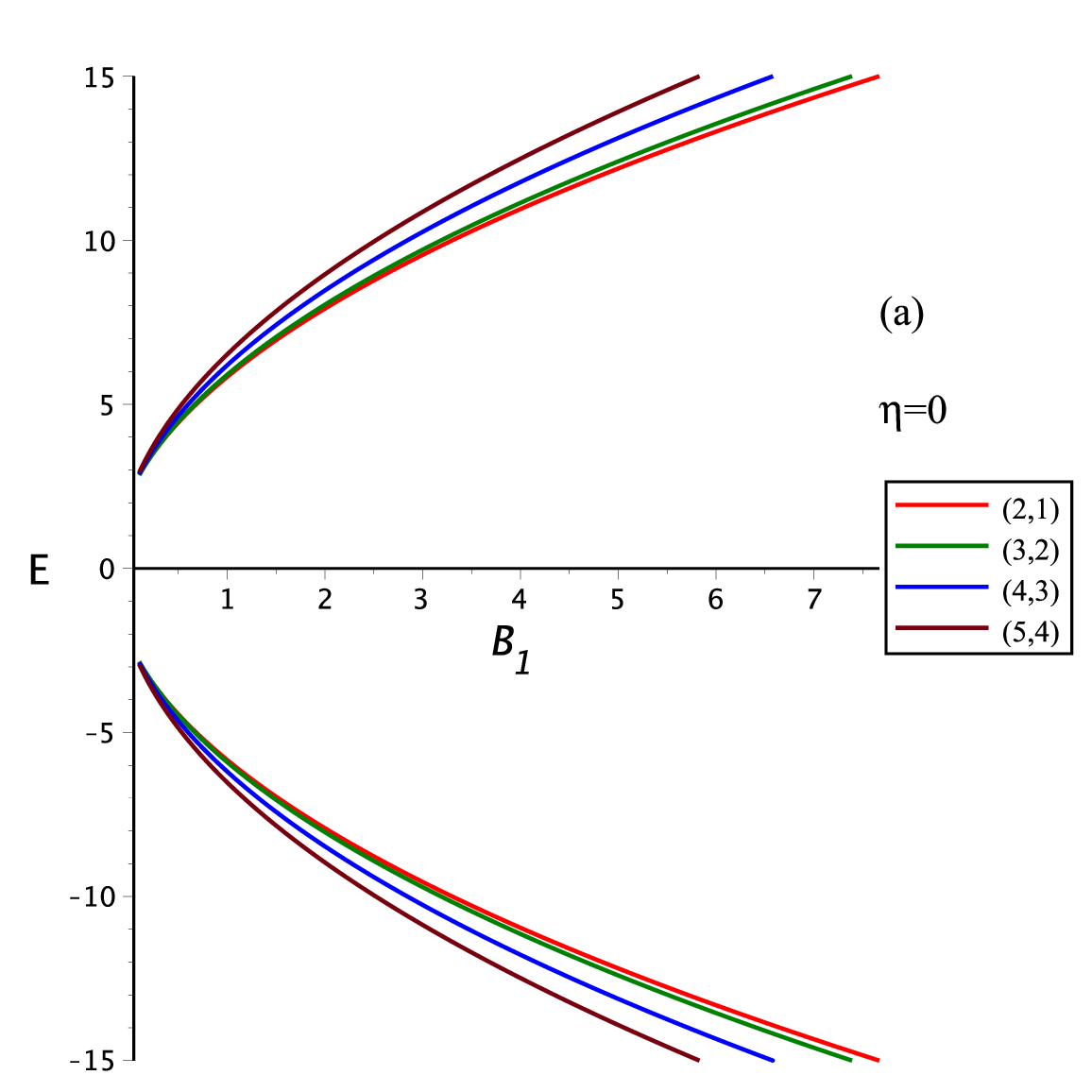}
\includegraphics[width=0.35\textwidth]{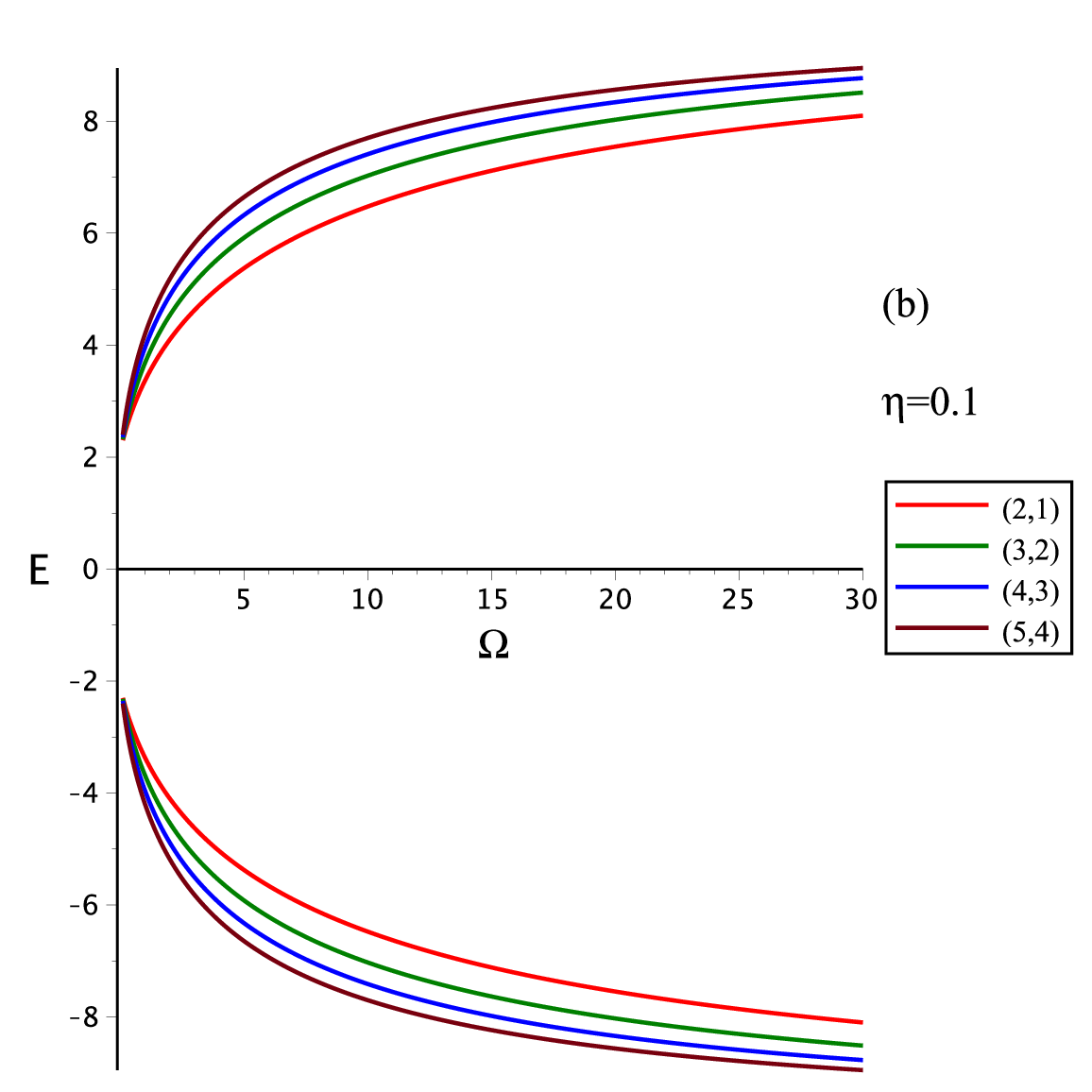} 
\includegraphics[width=0.35\textwidth]{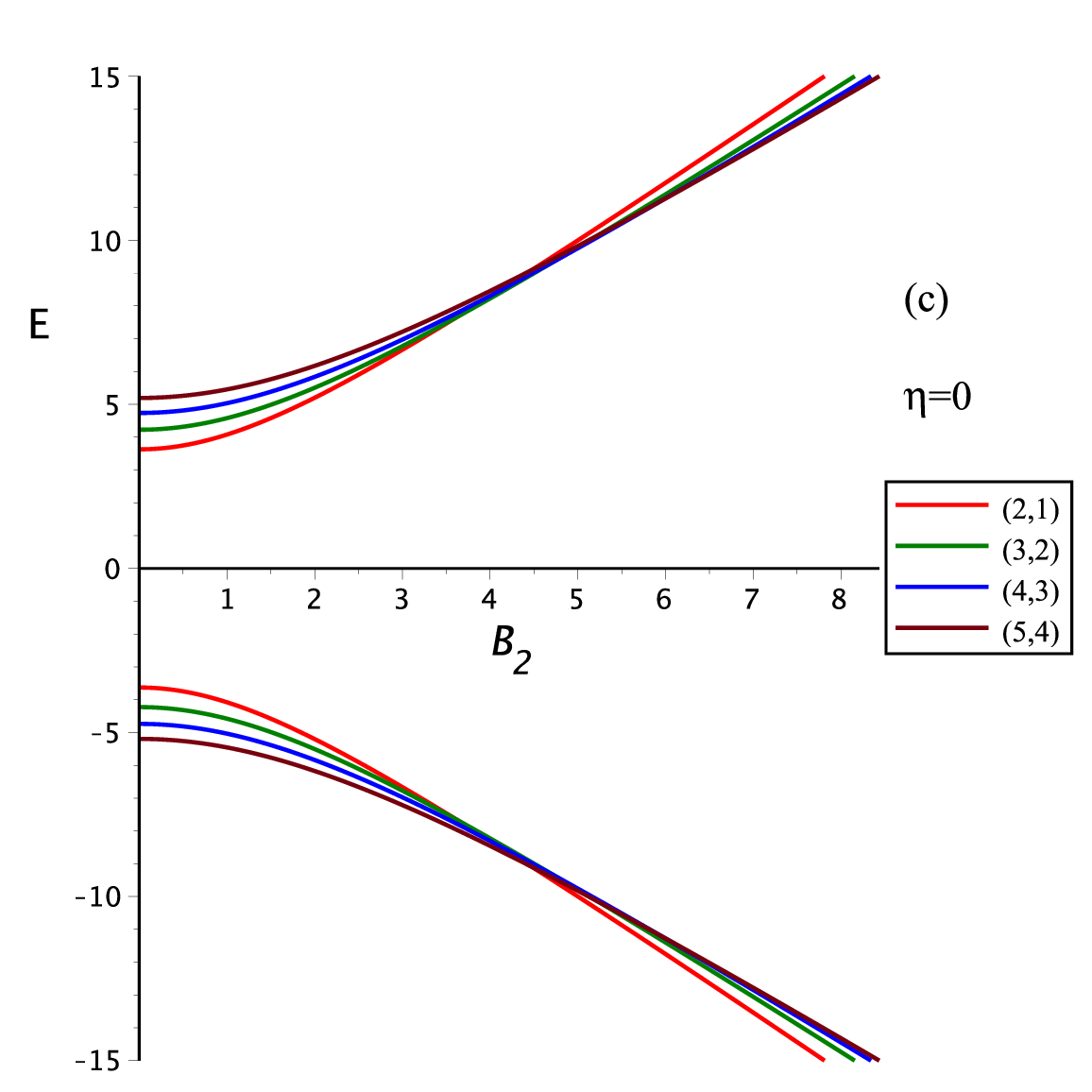}
\includegraphics[width=0.35\textwidth]{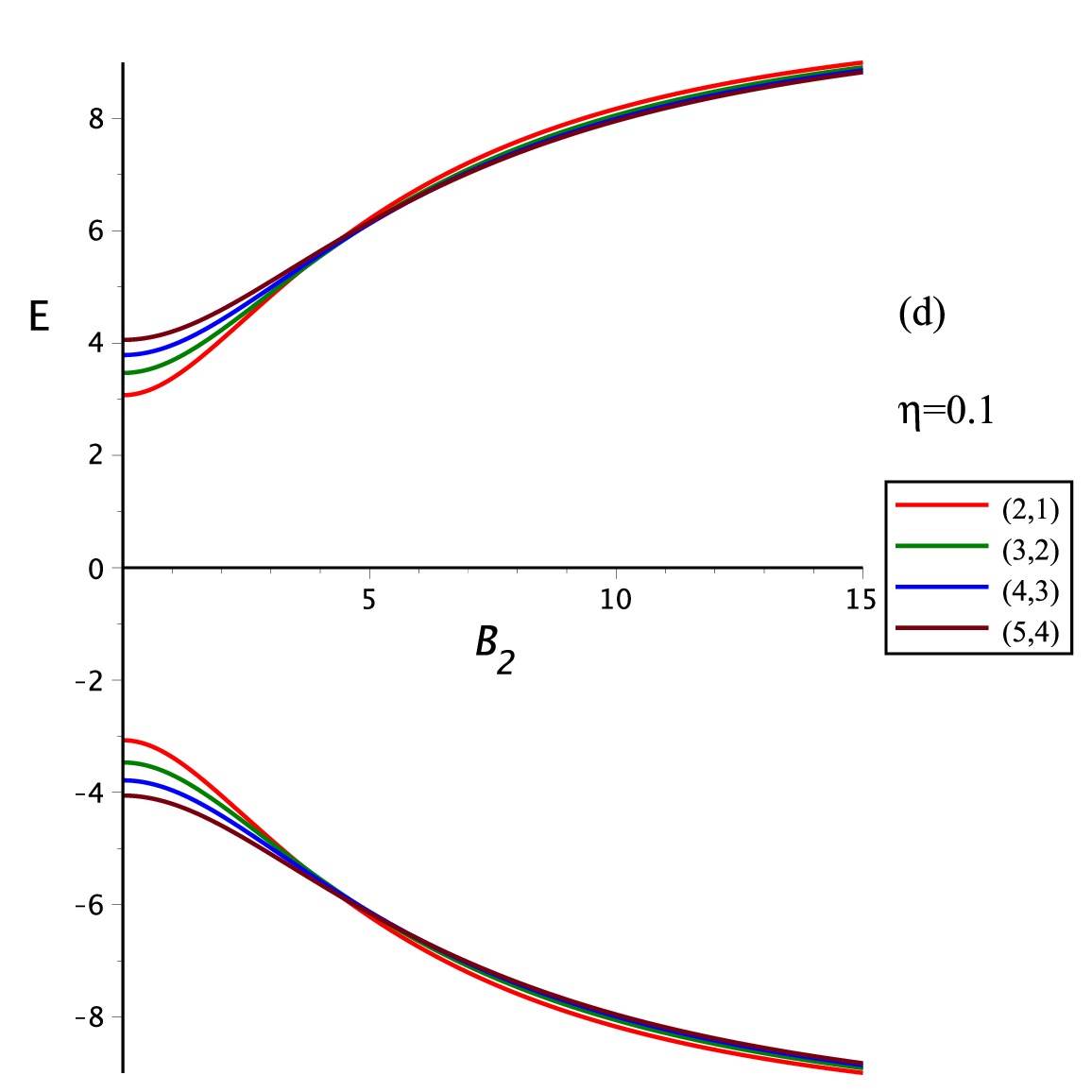}
\caption{\small 
{ The energy levels for the KG-oscillators in cosmic string rainbow gravity and a mixed magnetic field with the rainbow functions pair $g_0(y)=1,\, g_1(y)=\sqrt{1-\eta |E|}$, given by  Eq.(\ref{b27}),  plotted at $\alpha=0.5,\, m_\circ=1=q=k$  for different $(n_r,m)$-states, where (a) $E$ against $B_1$ for $\eta=0$ (no rainbow gravity) and $B_2=1$, (b) $E$ against the KG-oscillator frequency $\Omega$ for $\eta=0.1$ and $B_2=1$, (c) $E$ against $B_2$ for $\eta=0$ (no rainbow gravity) and $\Omega=1$, and (d) $E$ against $B_2$ for $\eta=0.1$ and $\Omega=1$.}}
\label{fig2}
\end{figure}%
\begin{gather}
  A_j\,\left[\tilde{\Omega}^3(j+\nu)-\frac{P_1}{2}\right] 
+A_{j+1}\left[\frac{\tilde{B}_1\tilde{B}_2 \tilde{\Omega}}{2}\left( j+\nu +1\right) -\frac{P_2}{2}\right] \notag \\
+A_{j+2}\left[ \frac{P_3}{2}-\frac{\tilde{\Omega}^2}{2}\left( j+\nu +1\right)
\left( j+\nu +2\right)\right]=0. \label{b18}
\end{gather}%
Consequently, one obtains, for $j=0,1,2,\cdots$, %
\begin{gather}
 A_{2}=\frac{A_0\,\left[\tilde{\Omega}^3(\nu)-\frac{P_1}{2}\right] 
+A_{1}\left[\frac{\tilde{B}_1\tilde{B}_2 \tilde{\Omega}}{2}\left( \nu +1\right) -\frac{P_2}{2}\right]}{\left[ \frac{\tilde{\Omega}^2}{2}\left( \nu +1\right)
\left( \nu +2\right)-\frac{P_3}{2}\right]},
\label{b19}
\end{gather}%
\begin{gather}
 A_{3}=\frac{A_1\,\left[\tilde{\Omega}^3(\nu+1)-\frac{P_1}{2}\right] 
+A_{2}\left[\frac{\tilde{B}_1\tilde{B}_2 \tilde{\Omega}}{2}\left( \nu +2\right) -\frac{P_2}{2}\right]}{\left[ \frac{\tilde{\Omega}^2}{2}\left( \nu +2\right)
\left( \nu +3\right)-\frac{P_3}{2}\right]},
\label{b20}
\end{gather}%
\begin{figure}[!ht]  
\centering
\includegraphics[width=0.3\textwidth]{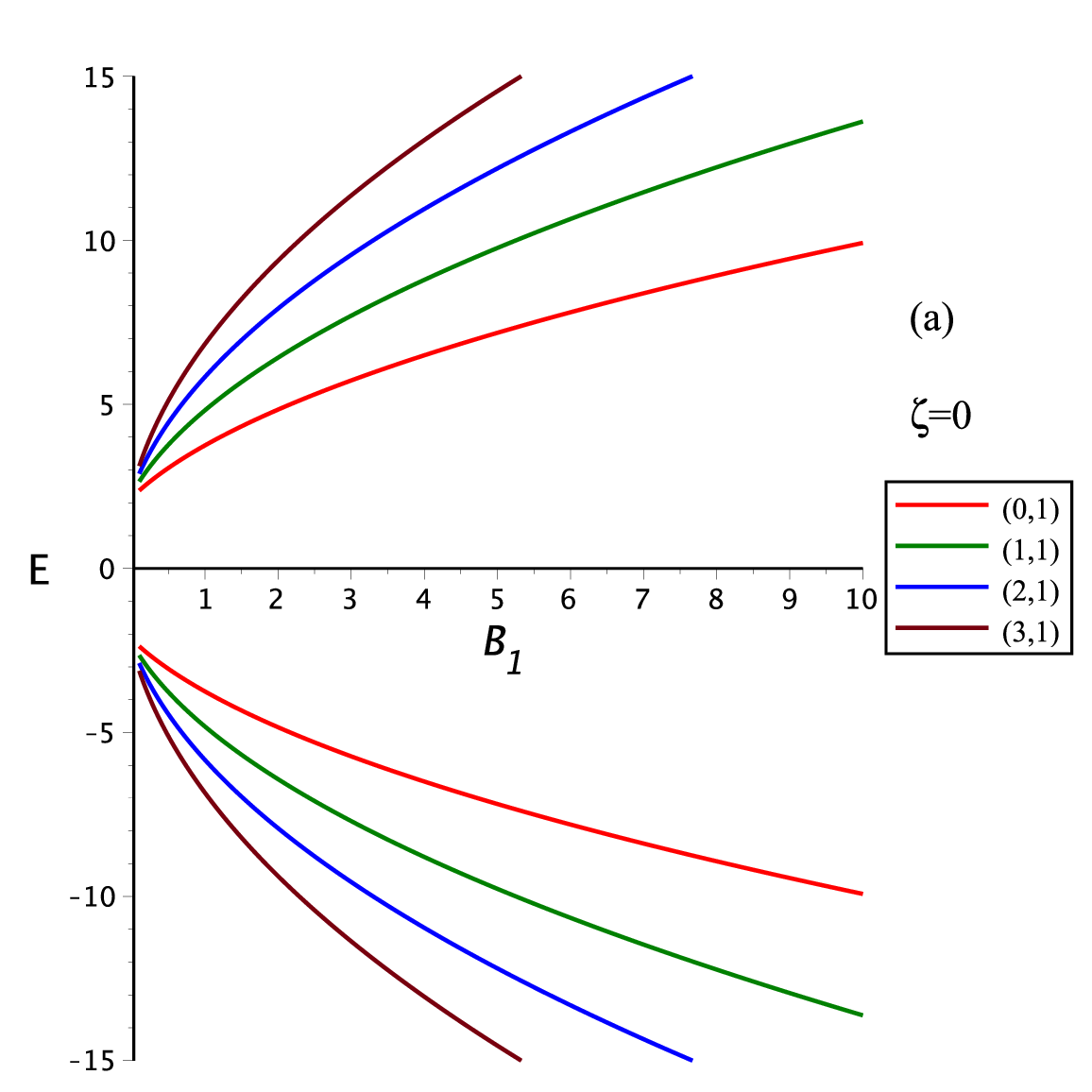}
\includegraphics[width=0.3\textwidth]{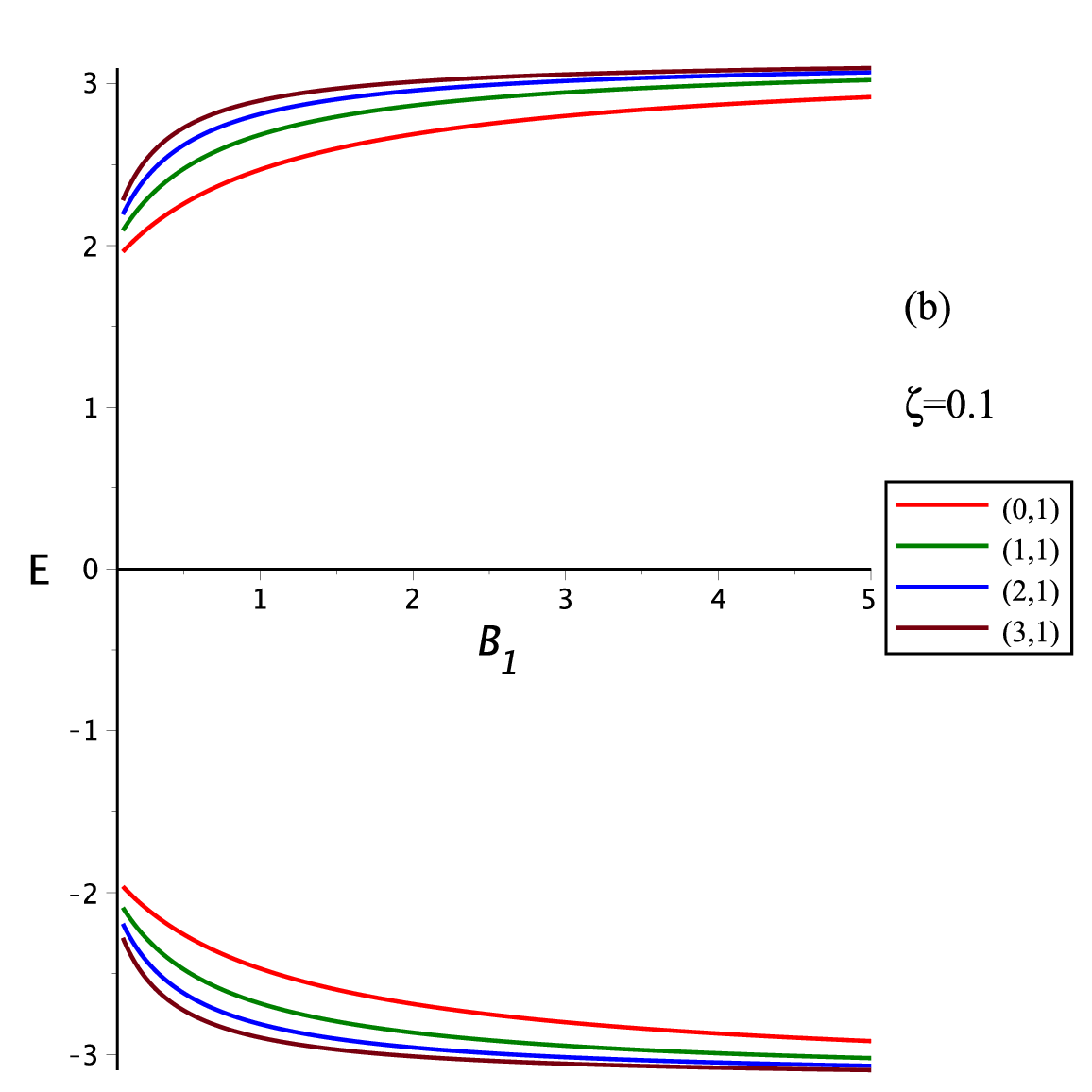} 
\includegraphics[width=0.3\textwidth]{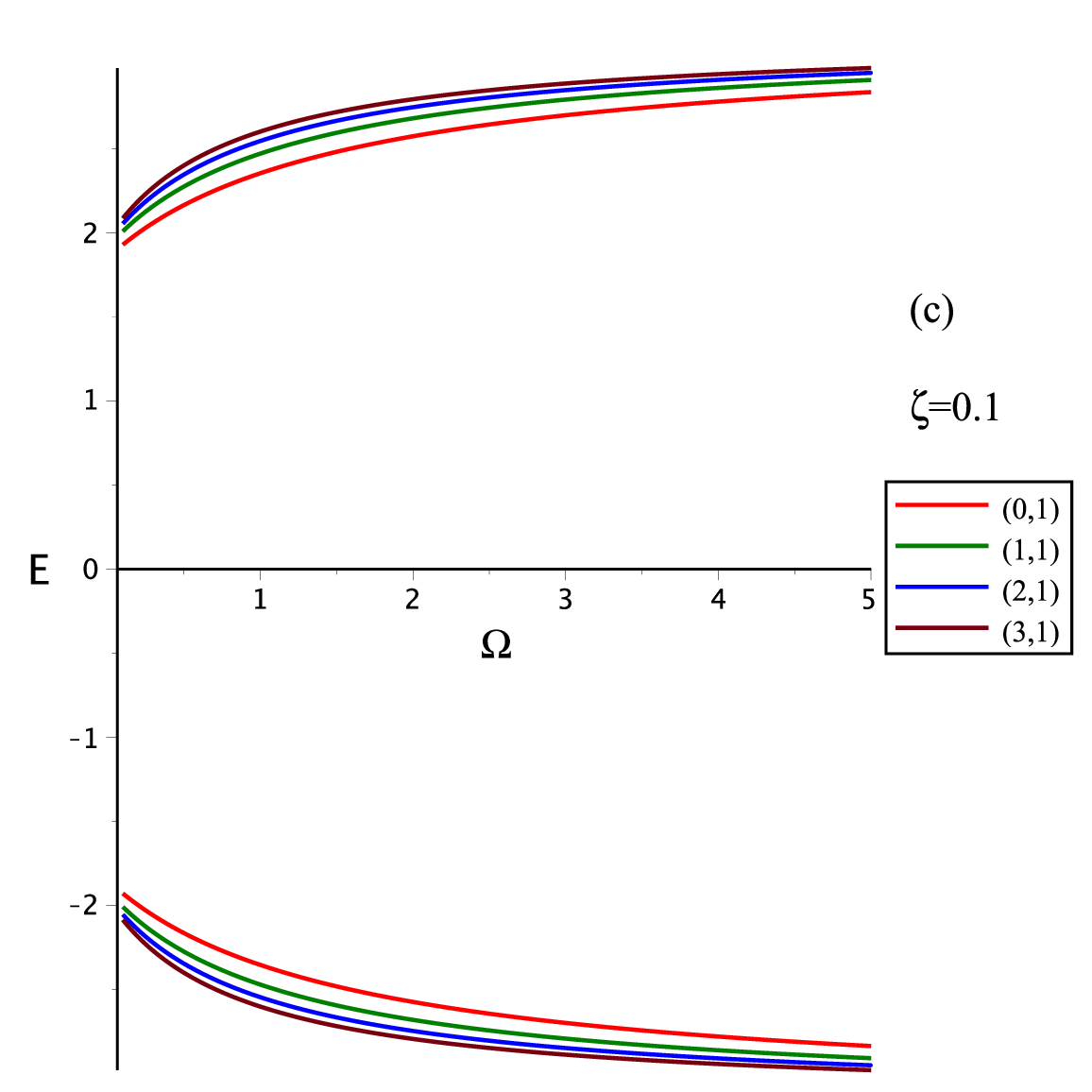}
\caption{\small 
{ The energy levels for the KG-oscillators in cosmic string rainbow gravity and a mixed magnetic field with the rainbow functions pair $g_0(y)=1,\, g_1(y)=\sqrt{1-\eta E^2}$, given by  Eq.(\ref{b28}),  plotted at $\alpha=0.5,\, m_\circ=1=q=k$  for $n_r=0,1,2,3,$ and $m=1$, where (a) $E$ against $B_1$ for $\eta=0$ (no rainbow gravity) and $B_2=1$, (b) $E$ against $B_1$ for $\eta=0.1$ and $B_2=1$, and (c) $E$ against the KG-oscillator frequency $\Omega$ for $\eta=0.1$ and $B_2=1$.}}
\label{fig3}
\end{figure}%
and so on. At this point, we suggest a new truncation condition that facilitates conditional exact solvability of the problem at hand and, at the same time, retrieves the result in (\ref{b10}). 

Our three terms recursion relation (\ref{b18}) allows us to use a valid/viable truncation condition so that $\forall{j}=n_r$ we may assume that $A_{n_r+2}=0$, $A_{n_r+1}\neq0$, and $A_{n_r}\neq0$. The first, $A_{n_r+2}=0$, would truncate the power series to a polynomial of order $(n_r+1)\geq1$. The second and the third would, in turn, validate the assumption that their coefficients identically vanish. That is,  $\forall{j}=n_r$, we have%
\begin{equation}
\frac{\tilde{B}_1\tilde{B}_2 \tilde{\Omega}}{2}\left( n_r+\nu +1\right) -\frac{P_2}{2}=0 \Rightarrow \Omega^2=\frac{\tilde{B}_1^2}{4\tilde{m}^2}\left[\left(n_r+|\tilde{m}|+\frac{3}{2}\right)^2-\tilde{m}^2\right], \label{b21}
\end{equation}
and%
\begin{equation}
\tilde{\Omega}^3(n_r+\nu)-\frac{P_1}{2}=0 \Rightarrow \lambda^2=2 \tilde{\Omega} (n_r+|\tilde{m}|+1)-\frac{\tilde{B}_1^2\tilde{B}_2^2}{4\tilde{\Omega}^2}. \label{b22}
\end{equation}%
Notably, while the result in (\ref{b22}) is consistent with that in (\ref{b10}), the result in (\ref{b21}) provides a parametric correlation between the magnetic fields $B_{1,z}={g_1(y)^2B_1/\alpha}$ and the KG-oscillator's frequency $\Omega$.  This correlation is clearly mandatory (and is an alternative to the condition provided by Ronveaux \cite{CR34}) for the condition used to obtain the result in (\ref{b10}).  Our result in (\ref{b22}) would consequently yield%
\begin{equation}
g_0(y)^2 E^2-m_\circ^2=g_1(y)^2\mathcal{G}_{n_r,m}, \label{b23}
\end{equation}%
where%
\begin{equation}
\mathcal{G}_{n_r,m}=2 \tilde{\Omega} (n_r+|\tilde{m}|+1)-\frac{\tilde{B}_1^2\tilde{B}_2^2}{4\tilde{\Omega}^2}+k^2+(\tilde{B}_2^2-\tilde{m}\tilde{B}_1)+2\Omega. \label{b24}
\end{equation}%

We start with the rainbow functions pair $g_0(y)=1,\,\, g_1(y)=\sqrt{1-\epsilon y}=\sqrt{1-\eta |E|}$,  where $\eta=\epsilon/E_p$.  In this case, our result in (\ref{b23}) would imply%
\begin{equation}
    E^2+\eta \, \mathcal{G}_{n_r,m}\,|E|-\left(m_\circ^2+\mathcal{G}_{n_r,m}\right)=0. \label{b25}
\end{equation}%
\begin{figure}[!ht]  
\centering
\includegraphics[width=0.35\textwidth]{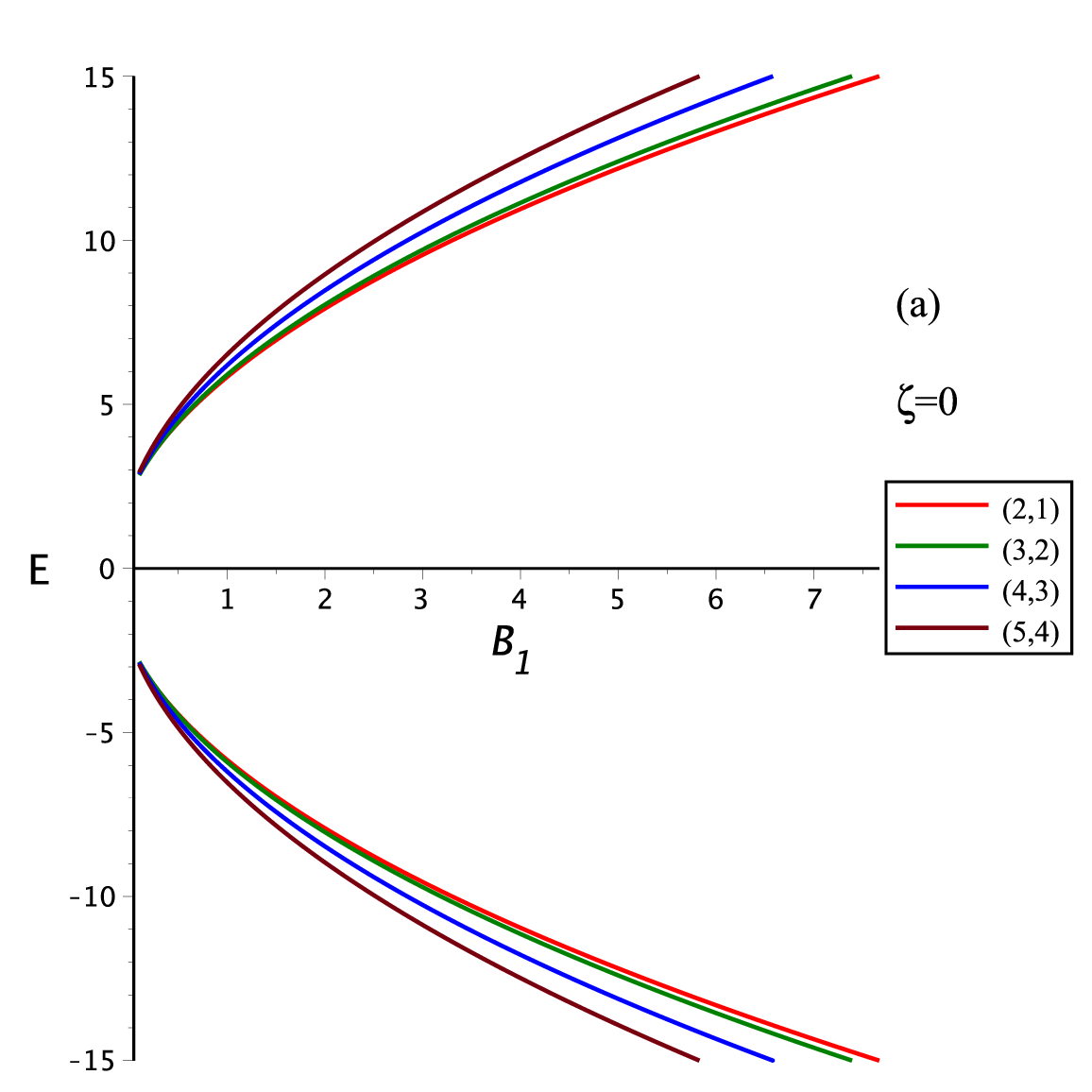}
\includegraphics[width=0.35\textwidth]{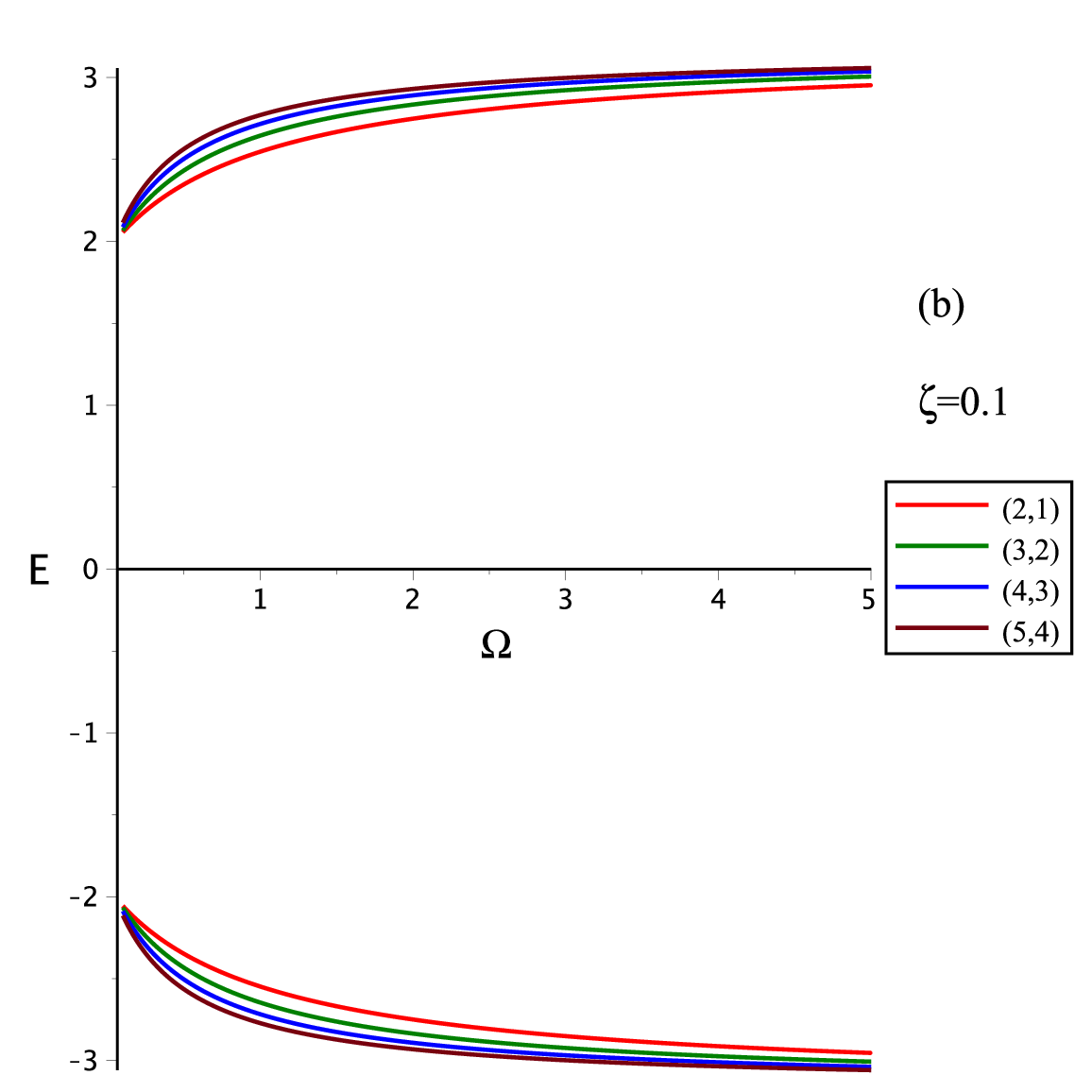} 
\includegraphics[width=0.35\textwidth]{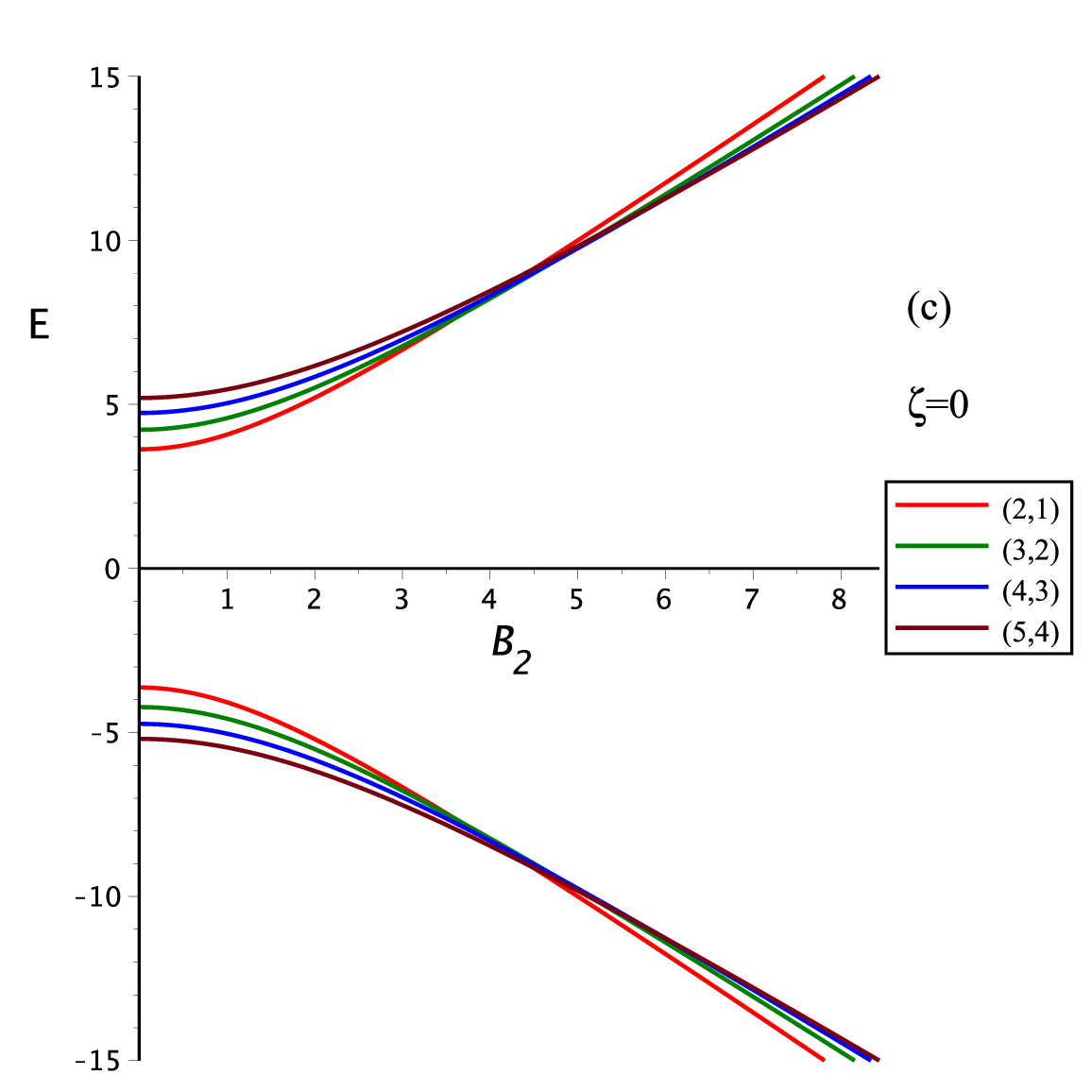}
\includegraphics[width=0.35\textwidth]{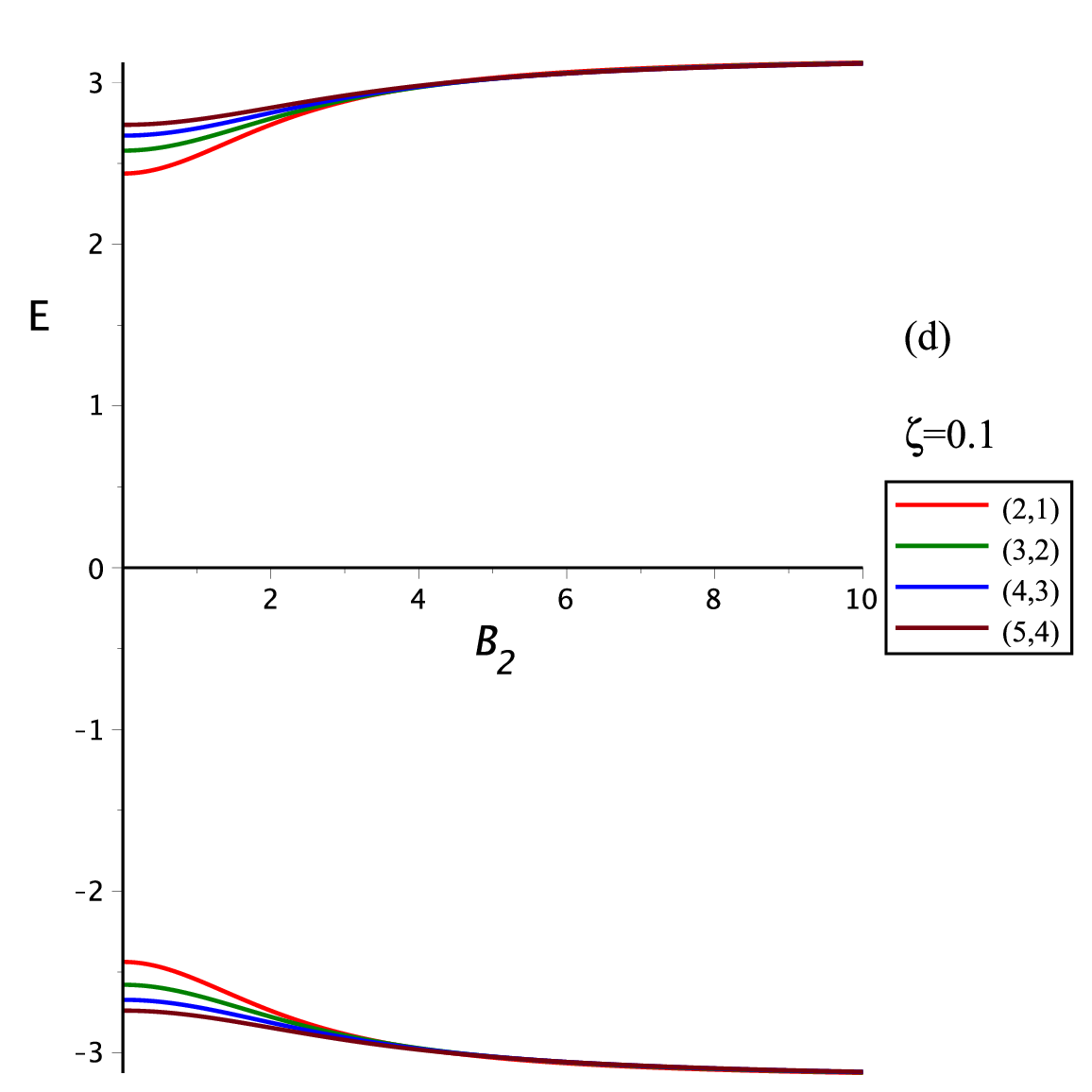}
\caption{\small 
{ The energy levels for the KG-oscillators in cosmic string rainbow gravity and a mixed magnetic field with the rainbow functions pair $g_0(y)=1,\, g_1(y)=\sqrt{1-\eta |E|}$, given by  Eq.(\ref{b28}),  plotted at $\alpha=0.5,\, m_\circ=1=q=k$  for different $(n_r,m)$-states, where (a) $E$ against $B_1$ for $\eta=0$ (no rainbow gravity) and $B_2=1$, (b) $E$ against the KG-oscillator frequency $\Omega$ for $\eta=0.1$ and $B_2=1$, (c) $E$ against $B_2$ for $\eta=0$ (no rainbow gravity) and $\Omega=1$, and (d) $E$ against $B_2$ for $\eta=0.1$ and $\Omega=1$.}}
\label{fig4}
\end{figure}%
Which would read%
\begin{equation}
E_+^2+\eta\, \mathcal{G}_{n_r,m}\, E_+ -\left(m_\circ+\mathcal{G}_{n_r,m}\right)=0 \Rightarrow E_+=-\frac{\eta \, \mathcal{G}_{n_r,m}}{2}+\sqrt{\frac{\eta^2 \, \mathcal{G}_{n_r,m}^2}{4}+\left(m_\circ^2+\mathcal{G}_{n_r,m}\right)}, \label{b26}
\end{equation}%
for $E=+E_+=+|E|$, and%
\begin{equation}
E_-^2-\eta\, \mathcal{G}_{n_r,m}\, E_- -(m_\circ+\mathcal{G}_{n_r,m})=0\Rightarrow E_-= +\frac{\eta \, \mathcal{G}_{n_r,m}}{2}-\sqrt{\frac{\eta^2 \, \mathcal{G}_{n_r,m}^2}{4}+(m_\circ^2+\mathcal{G}_{n_r,m})}, \label{b261}
\end{equation}%
for  $E=E_-=-|E|$. This would eventually yield%
\begin{equation}
 E_\pm=\mp\frac{\eta \, \mathcal{G}_{n_r,m}}{2}\pm\sqrt{\frac{\eta^2 \, \mathcal{G}_{n_r,m}^2}{4}+\left(m_\circ^2+\mathcal{G}_{n_r,m}\right)}. \label{b27}   
\end{equation}%
In Figures 1, we plot the energy levels for the KG-oscillators given by  Eq.(\ref{b27}),  for $(n_r,m)$-states with $n_r=0,1,2,3,$ and magnetic quantum number $m=1$ (i.e., the minimum allowed $m$ value as mandated by the correlation in (\ref{b21})). To observe rainbow gravity effect, we compare the energy levels, in 1(a), $E$ against $B_1$ at $\eta=0$ (no rainbow gravity) and $B_2=1$, with those in 1(b) for $\eta=0.1$. Moreover, the rainbow gravity effect is also observed in 1(c)  for $E$ against the KG-oscillator frequency $\Omega$ with $\eta=0.1$.  In Figure 2, we plot the energy levels for different $(n_r,m)$-states, namely, states labeled $(2,1), (3,2), (4,3), (5,4)$.  Where, the comparison between 2(a), with no rainbow gravity, and 2(b), with rainbow gravity, would clearly identify the rainbow gravity effect. In 2(b), without rainbow gravity, and 2(c), with rainbow gravity, we plot $E$ against $B_2$ and observe, along with rainbow gravity effect, energy levels crossings. Such energy levels crossings may very well be identified as occasional degeneracies due to  the structure of $\mathcal{G}_{n_r,m}$ in (\ref{b24}) and has nothings to do with rainbow gravity (as documented in 2(c)), but such degeneracies are rather manifestly introduced by the two $B_2$ competing terms  of $\mathcal{G}_{n_r,m}$ (i.e., the second and the fourth terms in (\ref{b24})). These energy levels crossing  are consequences of the effect of $B_2$, therefore. Such energy levels crossings turn the spectra of the KG-oscillator upside down, as $B_2$ grows up passing the crossing points. In both figures 1 and 2, we observe that under rainbow  gravity, the maximum allowed energy is given by $|E|_{max}=1/\eta=E_p$ for $\epsilon=1$. This is made obvious in the related figures as $|E|_{max}\approx 10$ for $\eta=0.1$ value used therein.

Next, we consider the rainbow functions pair $g_0(y)=1,\,\, g_1(y)=\sqrt{1-\epsilon y^2}=\sqrt{1-\zeta E^2}$,  where $\zeta=\epsilon/E_p^2$ in (\ref{b23}) to obtain%
\begin{equation}
E^2\,\left(1+\zeta\,\mathcal{G}_{n_r,m}\right)=m_\circ^2+\mathcal{G}_{n_r,m} \Rightarrow E=E_\pm=\pm\sqrt{\frac{m_\circ^2+\mathcal{G}_{n_r,m}}{\left(1+\zeta\,\mathcal{G}_{n_r,m}\right)}}. \label{b28}
\end{equation}%
In Figures 3 and 4, we plot the energy levels we plot the energy levels for (\ref{b28}) for $(n_r,m)$-states with $n_r=0,1,2,3,$ and magnetic quantum number $m=1$ (i.e., the minimum allowed $m$ value as mandated by the correlation in (\ref{b21})).  We observe the rainbow gravity effects through the comparison between the energy levels, in 3(a), against $B_1$ at $\eta=0$ (no rainbow gravity) and $B_2=1$, with those in 3(b) against $B_1$ and in 3(c) against the KG-oscillator frequency $\Omega$ for $\eta=0.1$. In Figure 4, we plot the energy levels for different $(n_r,m)$-states, namely, states labeled $(2,1), (3,2), (4,3), (5,4)$.  Where, the comparison between the energy levels in 4(a against $B_1$, in no rainbow gravity, and in 4(b) against $B_1$, in rainbow gravity, clearly documents the rainbow gravity effect. In 4(c) against $B_2$, without rainbow gravity, and 4(d) against $B_2$, with rainbow gravity, and again observe, along with rainbow gravity effect, energy levels crossings (such energy levels crossings, again, turn the spectra of the KG-oscillator upside down, in this case. ). Yet again in both figures 3 and 4, we observe that under, rainbow  gravity, the maximum allowed energy is given by $|E|_{max}=1/\sqrt{\zeta}=E_p$ for $\epsilon=1$. This is reflected on the related rainbow gravity figures 3 and 4 as $|E|_{max}\approx 3.16$ for $\zeta=0.1$ value used therein.

\section{Landau-like signature on the KG-oscillators energies }
Complementary to the above discussion, one may set $B_2=0$. In this case, the coefficients of $A_{n_r+1}$ would readily disappear and our correlation in (\ref{b21}) cease to exist (by the LHS part definition of the said correlation). As a result, our solution would be now classified as an exact solution with our $\mathcal{G}_{n_r,m}$ in (\ref{b24}) now reads
\begin{equation}
{\mathcal{G}}_{n_r,m}=2 \tilde{\Omega} (n_r+|\tilde{m}|+1)+k^2-\tilde{m}\tilde{B}_1+2\Omega. \label{c1}
\end{equation}%
Obviously, a Landau-like signature on the energy levels is introduced by the third term, $\tilde{m}\tilde{B}_1$.  Our energy levels reported in (\ref{b27}) and (\ref{b28}) remain the same but with our new  $\mathcal{G}_{n_r,m}$ in (\ref{c1}). 
In the absence of rainbow gravity, for example, the KG-oscillators' Landau-like  energy levels are given by%
\begin{equation}
    E_\pm=\pm\sqrt{m_\circ^2+2 \tilde{\Omega} (n_r+|\tilde{m}|+1)+k^2-\tilde{m}\tilde{B}_1+2\Omega}. \label{c2}
\end{equation}%
Of course, such Landau-like signatures can also be traced in the spectra of the KG-oscillators discussed in the preceding section in the mixed magnetic field used. 

At this point, one should observe that the same result in (\ref{c2}) is also obtainable using the intimate relation between the biconfluent Heun function and the confluent hypergeometric functions $H_B(\alpha, 0, \gamma, 0, z)=\,_1F_1\left(\frac{1}{2}+\frac{\alpha}{4}-\frac{\gamma}{4},1+\frac{\alpha}{2},z^2\right)$ in (\ref{b9}) and the related truncation condition for the confluent hypergeometric function to a polynomial of order $n_r\geq 0$ (i.e., $\gamma=4n_r+2+\alpha$ so that our $\alpha=2|\tilde{m}|$, $\gamma=\lambda^2/\tilde{\Omega}$, and $z=\sqrt{\tilde{\Omega}}r$ ).

\section{Concluding remarks}
In this paper, we have studied and  investigated KG-oscillators in cosmic string rainbow gravity spacetime in a mixed magnetic field generated by the 4-vector potential $A_\varphi=A_{1,\varphi}+A_{2,\varphi}=B_1 r^2/2+B_2 r$.  Such a 4-vector potential structure would introduce a magnetic field $\textbf{B}=\textbf{B}_1+\textbf{B}_2=\hat{z}\, B_z$, representing the superposition of two magnetic fields such that $A_{1,\varphi}\Rightarrow \textbf{B}_1=\hat{z}\,{g_1(y)^2}B_1/{\alpha}$ and $A_{2,\varphi}\Rightarrow \textbf{B}_2=\hat{z}\,{g_1(y)^2}B_2/{\alpha r}$ to yield the total magnetic field given in (\ref{a32}). Obviously, in no rainbow gravity (i.e., $g_1(y)=1$), $\textbf{B}_1$ is considered to be a uniform magnetic field, whereas $\textbf{B}_2$ is considered as a non-uniform one. Hence, the superposition of the two would be classified as non-uniform. At this point, the reader should be reminded that $\textbf{B}_1$ satisfies the homogeneous Maxwell equations whereas $\textbf{B}_2$ satisfies the inhomogeneous Maxwell equations in the non-null electrovacuum.  The later could only be generated by a non-zero current $J_\varphi=g_{\varphi\varphi} J^\varphi=-B_2\,g_1(y)^2/\mu_\circ r$  manifestly introduced by a string
at $r=0$, at which the current is divergent. Both magnetic fields, moreover, satisfy the two fundamental invariants of the electromagnetic fields (as discussed in more details in \cite{CR33}). Under rainbow gravity, however, the two magnetic fields (each at a time or both at the same time) become probe particle/anti-particle energy-dependent magnetic fields. Interestingly,  rainbow gravity does not only affect the spectroscopic structure of the quantum mechanical system at hand but also affects the magnetic field structure. 

The mixed magnetic field setting is used for the first time in the current proposal, to the best of our knowledge. Therefore, in order to clearly distinguish between the effects of rainbow gravity and the effects of the mixed magnetic field, we sought some rainbow functions pairs that fully comply with the intended rainbow gravity characteristic (i.e., $|E_{max}|\leq E_p$ ), for both particles and anti-particles alike. Our experience in \cite{CR28,CR29} has left us with no doubt that the loop quantum gravity motivated \cite{CR30,CR31} rainbow functions, (\ref{a31}), are the most suitable ones to use hereinabove. Only under such a sample model one may do healthful analysis, in our opinion. One may, nevertheless, wish to investigate, within the current methodical proposal setting, the rainbow function pair  $g_0(y)=(e^{\epsilon y}-1)/\epsilon y$, $g_1(y)=1$ associated with gamma-ray bursts \cite{CR13}. and/or use the the pair  $g_0(y)=g_1(y)=(1-\epsilon y)^{-1}$ associated with the horizon problem \cite{CR22,CR32}. The use of which already lies far beyond the scope of the current study.

In the light of our experience above,  the energy levels of the KG-oscillators in cosmic string rainbow gravity are observed to fully comply with rainbow gravity and secure their maximum possible energies to remain below the Planck energy scale, i.e., $|E_\pm|\leq E_p$, for the loop quantum gravity motivated rainbow functions in (\ref{a31}). Such rainbow gravity effect is readily documented in the figures reported above. 
\begin{figure}[!ht]  
\centering
\includegraphics[width=0.35\textwidth]{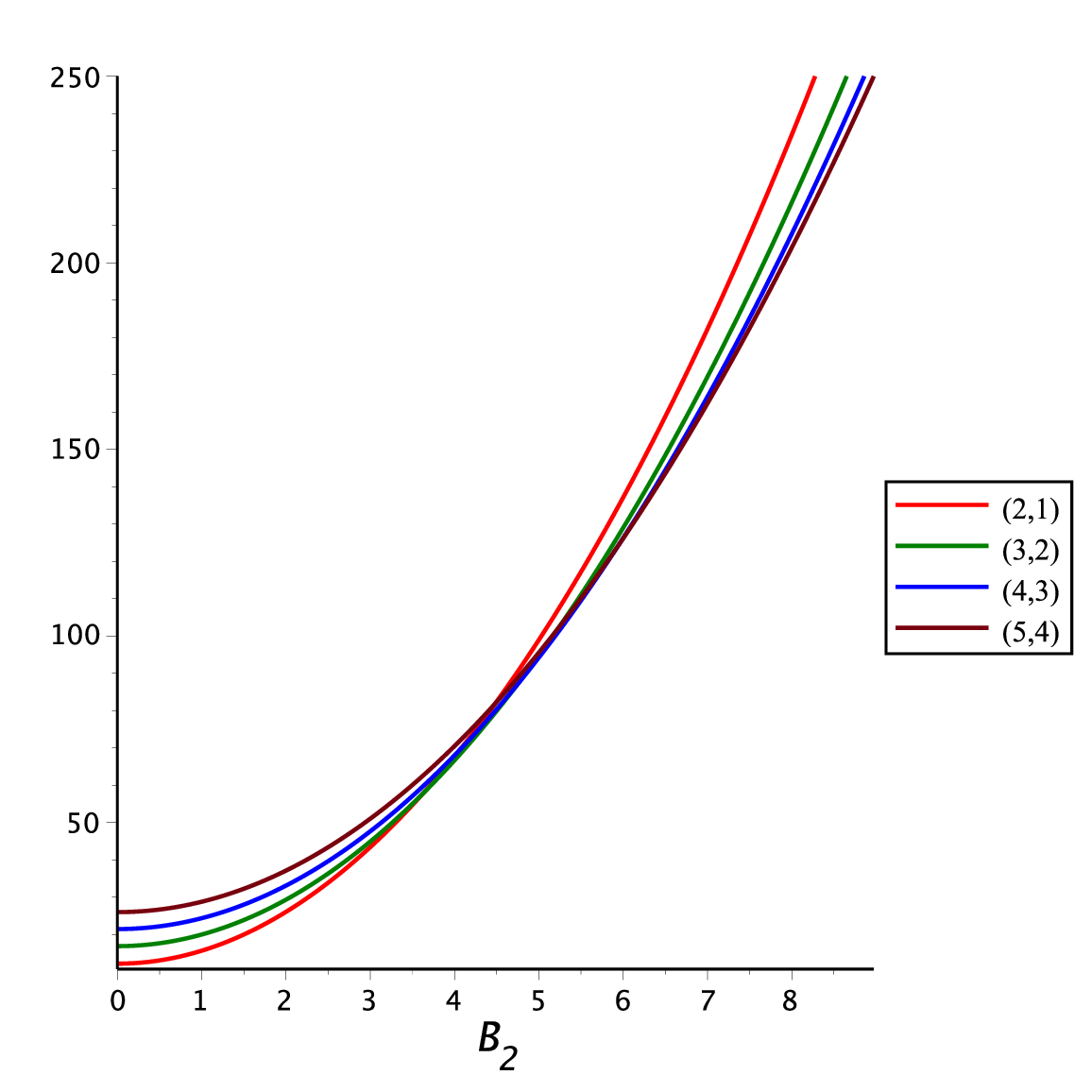}
\caption{\small 
{ Equation (\ref{d01}) for different values of $B_2$ using the corresponding values used in figure 2(c)}}
\label{fig5}
\end{figure}%
Notably, the correlation  (\ref{b21}) that facilitates the conditional exact solvability constraints our magnetic quantum number to satisfy $m\geq1$.
The quantum states with $m=0$ are left unfortunate, therefore. This is the price one has to pay in the absence of exact solvability of quantum mechanical systems, However, we have reported a significant set of quantum states that allowed us to adequately study/analyse the effects of the gravitational field of the cosmic string on the KG-oscillators under rainbow gravity settings and in a mixed magnetic field.  The most intriguing observation of the current study, nevertheless, are the energy levels crossings (or may very well be identified as occasional degeneracies) that are manifested in the process of increasing the magnetic field strength $B_2$. The energy levels crossings have flipped over the spectra in such away that the KG-oscillators' spectra is turned upside down. This effect is in fact a consequence of the structure of $\mathcal{G}_{n_r,m}$ in (\ref{b24})  (and has nothings to do with rainbow gravity as documented in 2(c)).
That is, in terms of $\tilde{B_1}$ and $\tilde{B_2}$ one may rewrite $\mathcal{G}_{n_r,m}$ of  (\ref{b24})), using the correlation (\ref{b21}) , as%
\begin{gather}
 \mathcal{G}_{n_r,m}=\frac{\tilde{B_1}}{|\tilde{m}|} (n_r+|\tilde{m}|+1) (n_r+|\tilde{m}|+\frac{3}{2})-\frac{\tilde{m}^2\tilde{B}_2^2}{(n_r+|\tilde{m}|+\frac{3}{2})^2}  \notag \\
 +\tilde{B}_2^2-\tilde{m}\tilde{B}_1+\frac{\tilde{B_1}}{|\tilde{m}|}\sqrt{(n_r+|\tilde{m}|+\frac{3}{2})^2-\tilde{m}^2}+ k^2,   \label{d01}
\end{gather}%
where, by the correlation (\ref{b21}), we have used %
\begin{equation}
\tilde{B_1}=\frac{2|\tilde{m}|\Omega}{\sqrt{(n_r+|\tilde{m}|+\frac{3}{2})^2-\tilde{m}^2}}. \label{d02}
\end{equation}
Therefore, the implicit competition between the two magnetic fields in (\ref{d01}) shapes the spectroscopic structure and yields energy levels crossing that consequently turns the spectra upside down as $B_2$ increases.
In figure 5, we plot $\mathcal{G}_{n_r,m}$ of (\ref{d01}) against $B_2$ for the corresponding values used in figure 2(c). Figure 5 clearly indicates that the behavior of $\mathcal{G}_{n_r,m}$ against $B_2$ is inherited by the energy levels against $B_2$ plotted and reported above. One should clearly observe that the contribution of the second term in (\ref{d01}) decreases for larger quantum numbers while the third term remains the same, for a given $B_2$. 

Our new truncation approach of the biconfluent Heun function to a polynomial of order $(n_r+1)\geq1$ (and not to a polynomial of order $n_r\geq0$) has manifestly introduced a new alternative condition to those described by Ronveaux \cite{CR34} and/or by  Ishkhanyan et al \cite{CR35}. Such a new truncation recipe retrieves the results of Ronveaux \cite{CR34}  and/or  Ishkhanyan et al \cite{CR35} on the truncation of biconfluent Heun function $H_B(\alpha, \beta, \gamma, \delta, z)$ to a polynomial of order $n_r\geq 0$ (and not of order $(n_r+1)\geq0$) for $\gamma=2(n_r+1)+\alpha$. We have rather used the new truncation conditions so that $\forall{j}=n_r$ we may assume that $A_{n_r+2}=0$, $A_{n_r+1}\neq0$, and $A_{n_r}\neq0$.  Whilst the first, $A_{n_r+2}=0$, would truncate the power series into a polynomial of order $(n_r+1)\geq1$, the vanishing coefficients of the second, $A_{n_r+1}\neq0$, would yield that $\delta--\beta(2n_r+\alpha+3)$ as an alternative recipe (the details on such recipe of the solution of the biconfluent Heun equation are discussed in the Appendix below). 

Finally, our analysis above provides a brute force documentation that supports the argument of Bezerra et al \cite{CR17} on that rainbow gravity is not merely a mathematical time-coordinate rescaling. We have clearly observed that rainbow gravity not only significantly affects the spectroscopic structure of the quantum particles but also renders the magnetic field to be energy-dependent. To the best of our knowledge, KG-oscillators in a cosmic string rainbow gravity in the above  mixed magnetic field have never been discussed elsewhere. Yet, the current methodical proposal provides a set of conditionally exactly solvable models that are very frequently used to study the gravitational fields effects that are of interest in not only quantum gravity but also in condense matter physics.%

\section{Appendix: Biconfluent Heun equation conditional exact solvability}

In this appendix we would like to recollect the biconfluent Heun equation and introduce a new conditionally exact solution recipe. The biconfluent Heun functions $H_B(\alpha,\beta,\gamma,\delta,x)$ are known to be the solutions for the Heun equation \cite{CR34,CR35}
\begin{equation}
x y^{\prime\prime}(x)+(\tilde{\alpha}-\beta x -2 x^2)y^{\prime}(x)
+(\tilde{\gamma} x -\tilde{\delta}) y(x)=0, \label{A1}
\end{equation}%
where%
\begin{equation}
\tilde{\alpha}=1+\alpha, \,\, \tilde{\gamma}=\gamma-\alpha-2, \,\, \tilde{\delta}=\frac{1}{2}\left[\delta+(1+\alpha)\beta\right] \label{A2}
\end{equation}%
The power series expansion in the form of%
\begin{equation}
  y(x)=x^\nu \sum\limits_{j=0}^{\infty }A_j\, x^j,  \label{A3}
\end{equation}%
would, in a straightforward manner, imply%
\begin{gather}
\sum\limits_{j=0}^{\infty }\left\{ A_j\,\left[\tilde{\gamma}-2(j+\nu)\right] 
-A_{j+1}\left[\tilde{\delta}+\beta( j+\nu +1) \right] 
+A_{j+2}\left[ \left( j+\nu +2\right)
\left( \tilde{\alpha}+j+\nu +1\right)\right] \right\} x^{j+\nu +1} \notag \\
+\left\{ A_{1}\left[(\nu+1)(\tilde{\alpha}+\nu)\right] -A_{0}\left[ \tilde{\delta}+\beta \nu\right] \right\} x^{\nu} +A_{0}\left[\nu(\tilde{\alpha}+\nu-1 \right] x^{\nu-1}=0.
\label{A4}
\end{gather}%
The last term of which would suggest that since $A_0\neq 0$ we have $\nu=0$. Consequently, $A_1=A_0 \tilde{\delta}/\tilde{\alpha}$ and%
\begin{equation}
 A_j\,\left[\tilde{\gamma}-2j\right] 
-A_{j+1}\left[\tilde{\delta}+\beta ( j+1) \right] 
+A_{j+2}\left[ \left( j+2\right)
\left( \tilde{\alpha}+j+1\right)\right]=0.   \label{A5}
\end{equation}%
At this point. we impose the conditions that $\forall{j}=n_r$ we set $A_{n_r+2}=0$,  $A_{n_r+1}\neq0$ and $A_{n_r}\neq0$. The first condition, $A_{n_r+2}=0$, would truncate the power series to a polynomial of order $(n_r+1)\geq1$. Moreover, we may facilitate the so called conditional exact solvability by the requirement that the coefficients of $A_{n_r+1}\neq0$ are vanishing, i.e.,%
\begin{equation}
 \tilde{\delta}+\beta( n_r+1)=0; \, \forall j=n_r \Rightarrow \delta--\beta(2n_r+\alpha+3). \label{A6}    
\end{equation}%
Furthermore, $A_{n_r}\neq0$ implies $\tilde{\gamma}-2n_r=0 \Rightarrow \gamma=2(n_r+1)+\alpha$, which is in fact in exact accord with that reported by \cite{CR34} and/or by  Ishkhanyan et al \cite{CR35}, provided that the parametric correlation in (\ref{A6}) is satisfied. 

The above recipe may very well be used for the solution of  confluent Heun equation that has very recently been followed and successfully implemented in the study of  KG-oscillators in Eddington-inspired Born-Infeld gravity global monopole spacetime and a Wu-Yang magnetic monopole by Mustafa et al \cite{CR36}.
\

\textbf{Data availability statement:} 
The authors declare that the data supporting the findings of this study are available within the paper. 

\textbf{Declaration of interest:}
The authors declare that they have no known competing financial interests or personal relationships that could have appeared to influence the work reported in this paper.

\end{document}